\documentclass[3p,11pt,authoryear]{elsarticle}

\usepackage{amssymb}
\usepackage{mathrsfs}
\usepackage{rotating}
\usepackage{graphicx}
\usepackage{longtable,lscape}

\newcommand{\fn}[1]{\footnote{\scriptsize{#1}}} 

\newcommand{\Eqn}[1]{Eq{#1}.}  
\newcommand{\Fig}[1]{Fig{#1}.}  
\newcommand{\ud}{\mathrm{d}}  
\newcommand{\Cassit}{\textit{Cassini}}  
\newcommand{\Voyit}{\textit{Voyager}}  

\newcommand\araa{Ann.~Rev. Astron.~Astrophys.}%
\newcommand\aj{Astron.~J.}%
\newcommand\apj{Astrophys.~J.}%
\newcommand\apjl{Astrophys.~J.~Lett.}%
\newcommand\grl{Geophys.~Res.~Lett.}%
\newcommand\icarus{Icarus}%
\newcommand\jgr{J.~Geophys.~Res.}%
\newcommand\nat{Nature}%
\newcommand\planss{Planet. Space Sci.}%
\newcommand\ssr{Space Sci. Rev.}%

\journal{Icarus}

\begin{document} 

\begin{frontmatter}
\title{Probing the inner boundaries of Saturn's A~ring\\with the Iapetus~-1:0 nodal bending wave} 

\author{Matthew~S.~Tiscareno$^1$, Matthew~M.~Hedman$^1$, Joseph~A.~Burns$^{1,2}$,\\John~W.~Weiss$^{3,4}$, \& Carolyn~C.~Porco$^3$}

\address{$^1$Department of Astronomy, Cornell University, Ithaca, NY 14853, USA.\\$^2$College of Engineering, Cornell University, Ithaca, NY 14853, USA.\\$^3$CICLOPS, Space Science Institute, 4750 Walnut Street, Boulder, CO 80301, USA.\\$^4$Physics and Astronomy Department, Carleton College, 1 North College Street, Northfield, MN 55057, USA.}

\begin{abstract}
The Iapetus~-1:0 nodal bending wave, the first spiral wave ever described in Saturn's rings, has been seen again for the first time in 29 years.  We demonstrate that it is in fact the nodal bending wave, not the 1:0 apsidal density wave as previously reported.  We use wavelet analysis to determine the wavelength profile, thus deriving the surface density at every point in the region covered by the bending wave. This profile is consistent with surface densities measured from more localized spiral density waves in the outer Cassini Division and the inner and mid-A Ring, varying smoothly from the low values of the former to the higher values of the latter.  

Most remarkably, our analysis indicates that there is no significant change in surface density across the boundary between the outer Cassini Division and the inner-A~ring, despite the very abrupt increase in optical depth and reflected brightness at this location.  We consider anew the nature of the classically identified ``inner edge of the A~ring,'' given that it does not appear to be correlated with any abrupt increase in surface density.  There is an abrupt increase in surface density at the Pandora~5:4 density wave, $\sim$300~km outward of the A~ring's inner edge.  Further study is needed to robustly interpret our findings in terms of particle properties and abundances, much less to explain the origins of the implied structure. 

\end{abstract}


\end{frontmatter}

\section{Introduction}
Spiral density waves and spiral bending waves, generated in a disk by resonant interactions with perturbing moons, are a pervasive and useful feature in Saturn's rings \cite[see, e.g.,][]{Ringschapter13}.  The very first spiral wave ever reported in Saturn's rings, confirming the prediction of \citet{GT78b,GT79,GT80} that the spiral phenomena well-known in galaxies should also appear in planetary rings, was described by \citet{Cuzzi81} from \Voyit{~1} imaging data and identified as the Iapetus~1:0 apsidal density wave.  The long-wavelength Iapetus wave was the only one resolvable in the comparatively low-resolution \Voyit{~1} images, which were the first available rings data as release of the \Voyit{~1} RSS radio occultation data was delayed \citep{Tyler83} and the PPS stellar occultation was not carried out until \Voyit{~2} \citep{Espo83}. 

Eventually, a panoply of density and bending waves were observed in the \Voyit{} RSS and PPS data sets \citep[see references in][]{soirings}, but curiously, the Iapetus wave described by \citet{Cuzzi81} was never seen again by \Voyit{}.  Similarly, when \Cassit{} imaging and occultations revealed a new level of finer-scale spiral waves \citep[see references in][]{ColwellChapter09}, the Iapetus wave still was not seen.  

\begin{figure*}[!t]
\begin{center}
\includegraphics[width=11cm]{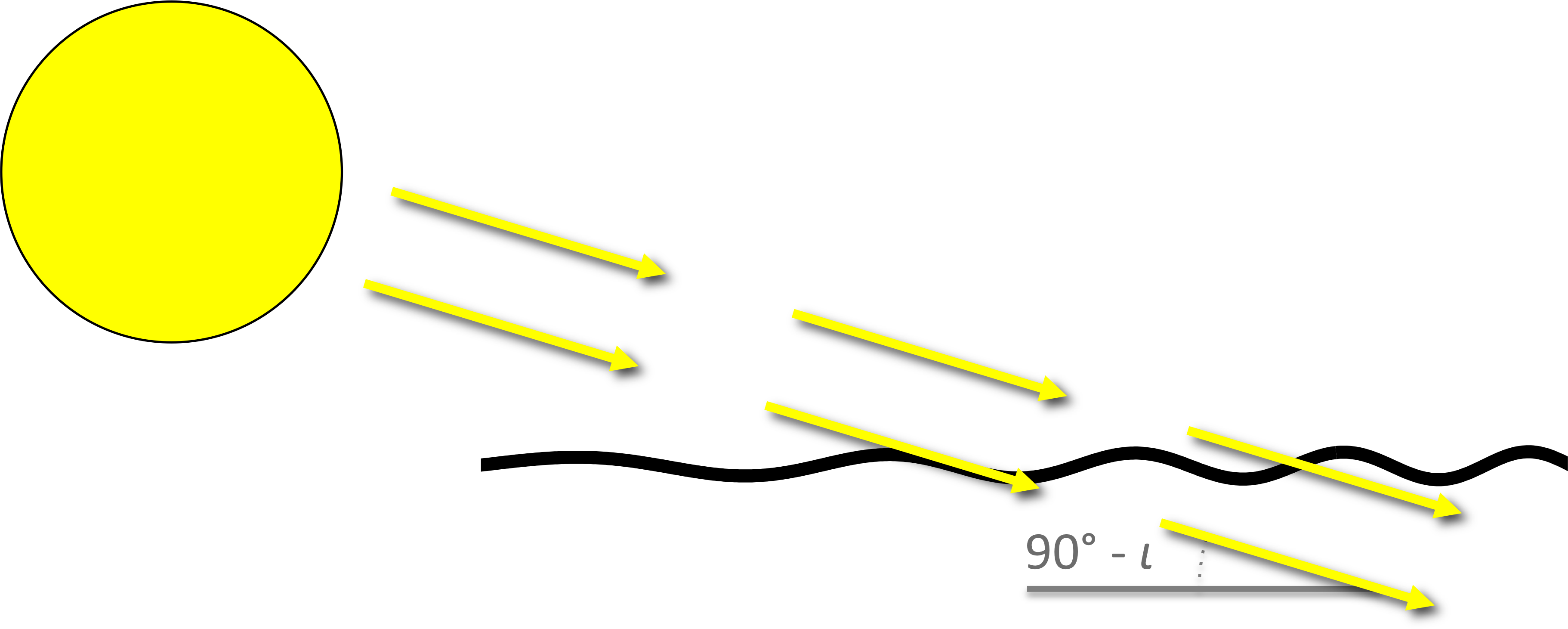}
\caption{In this cartoon, the corrugation of the ring by a bending wave, seen edge-on, is represented by the black line.  The yellow circle represents the Sun, and the yellow arrows represent the path of sunlight.  The complement of the solar incidence angle~$\iota$ is marked in gray.  When the Sun is low on the horizon (i.e., $\iota$ is near $90^\circ$) sunlight passes through at a shallower angle in some regions of the wave than in others (represented by the lower and upper ray paths, respectively), leading to a modulation in the measured optical depth that corresponds with the corrugations.  This effect is only seen near local noon or midnight; by contrast, at local morning and evening, sunlight passes along the peaks and troughs (into or out of the page, in this representation) rather than across them.  For purposes of readability, the wave amplitude is exaggerated in this cartoon; there is no evidence that slopes in the Iapetus nodal bending wave are ever as large as the angle of incident sunlight \citep[cf.][]{Gresh86,RL88}, even during equinox, when the angular size of the Sun enforces a minimum value for the latter.  \label{BendingWaveCartoon}}
\end{center}
\end{figure*}

We report here that the long-lost Iapetus wave was pointed out for the first time in 29 years in \Cassit{} images taken during 2009.  As when \Voyit{~1} discovered this wave, Saturn was close to equinox at the time the wave was recovered.  The \Voyit{~1} flyby of 1980~November~12 was eight months after the 1980~March~3 equinox, and similarly, the Iapetus wave is most prominent in images taken within about a year of the 2009~August~11 equinox. 
By contrast, it is very difficult to see in images taken closer to \Cassit{}'s 2004 arrival at Saturn, 
as well as in the most recent images from 2012.  
Around equinox, the edge-on approach of incident sunlight to the rings (\Fig{}~\ref{BendingWaveCartoon}) highlighted vertical structure of many kinds, including radial corrugations \citep{HedmanCorrugation11}, impact ejecta clouds \citep{Impactclouds13}, vertically scalloped gap edges \citep{Weiss09,SP10}, shadows cast by ``propellers'' and other compact embedded structures \citep{Giantprops10,SP10}, and shadows cast on the rings by moons.\fn{See images PIA11498, PIA11506, PIA11634, PIA11651, and PIA11660 at \texttt{http://photojournal.jpl.nasa.gov}}  The association with equinox, along with our detailed analysis of the wavetrain itself (see Section~\ref{Analysis}), confirm the long-standing suspicion that \citet{Cuzzi81} saw not the Iapetus~1:0 apsidal density wave, but rather the vertically corrugated Iapetus~-1:0 nodal bending wave (see Section~\ref{SpiralWaves}). 

After an overview of the spatial context of the wave in the next section, we give detailed comments on the nature of resonances and spiral waves in Section~\ref{SpiralWaves}.  Then we describe our observations in Section~\ref{Observations} and our wavelet analysis of the Iapetus nodal bending wave in Section~\ref{Analysis}.  Discussion of our results and their implications for the outer Cassini Division and inner~A ring is contained in Section~\ref{Discussion}.  

\begin{figure*}[!t]
\begin{center}
\hspace{0.65cm} \includegraphics[width=15.1cm]{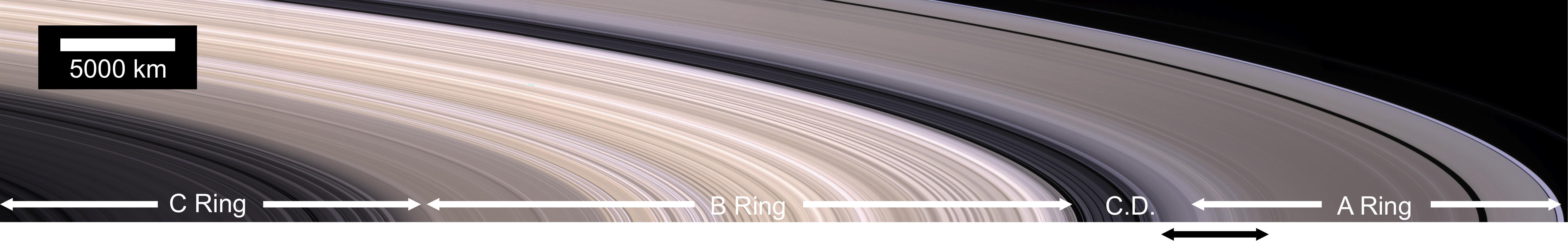}
\caption{An overview of Saturn's main rings, with the primary regions marked.  The Cassini Division (marked ``C.D.'') is between the A~and B~rings.  Along the bottom edge of the image, the black double-headed arrow indicates the region of interest for this work, which is shown in more detail in \Fig{}~\ref{ContextFig2}. 
\label{ContextFig1}}
\end{center}
\end{figure*}

\begin{figure*}[!t]
\begin{center}
\hspace{0.65cm} \includegraphics[width=15.1cm]{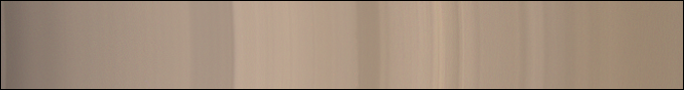}
\includegraphics[width=16cm]{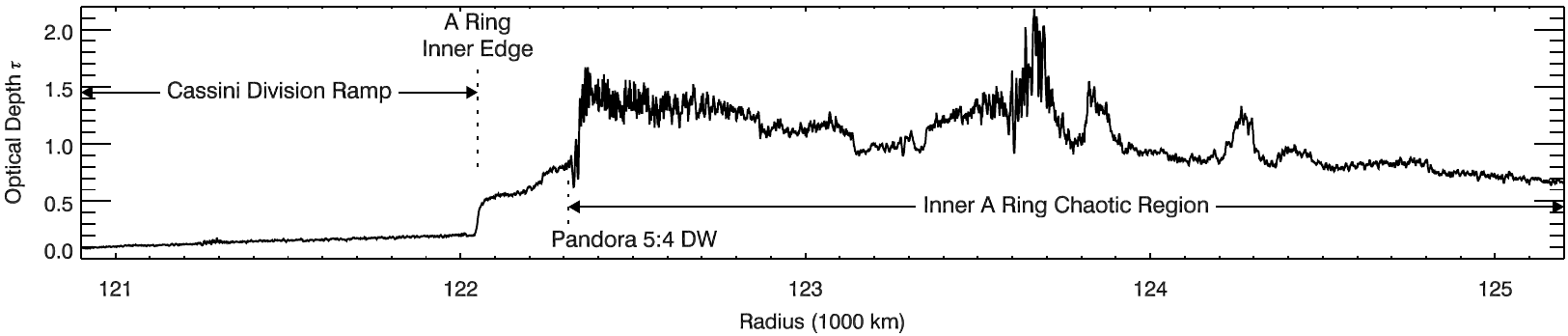}
\caption{The region of interest for this work (see \Fig{}~\ref{ContextFig1} for larger context) consists of the outer Cassini Division and the inner A~ring.  The bottom panel shows the optical depth $\tau$ for the region as measured by a \Cassit{}~VIMS stellar occultation (tracking the attenuation of the star $\gamma$~Crucis as \Cassit{} saw it pass behind the rings on 2008~October~16).  The top panel is an image mosaic taken 2008~November~26, showing a fairly typical (non-equinox) view of the region, with apparent brightness correlating with the optical depth structure.  
\label{ContextFig2}}
\end{center}
\end{figure*}

\begin{figure*}[!t]
\begin{center}
\hspace{1.15cm} \includegraphics[width=15.15cm]{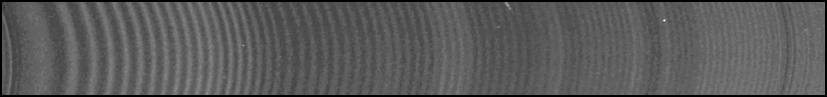}
\includegraphics[width=16.5cm]{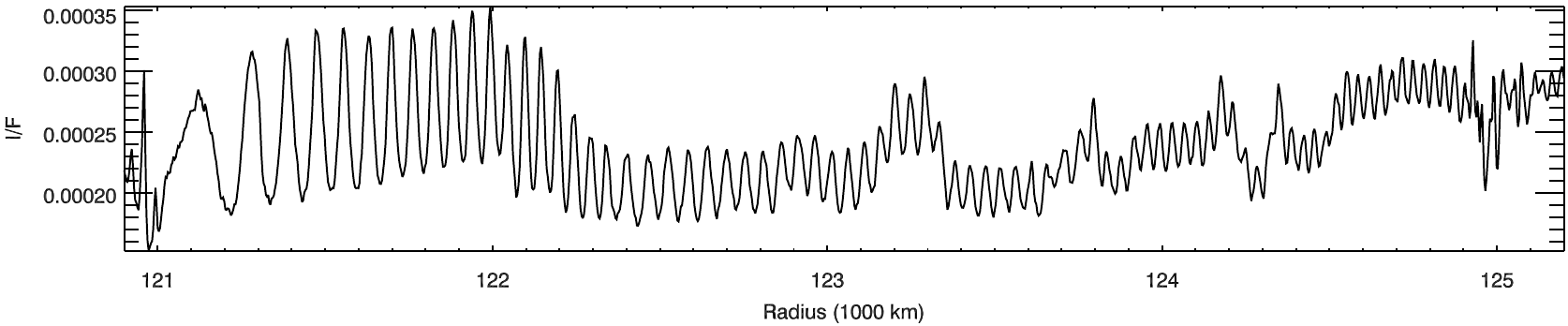}
\caption{The same region shown in \Fig{}~\ref{ContextFig2} as seen during equinox, with the optical depth structure greatly muted and the Iapetus nodal bending wave dominant.  The upper panel is an image mosaic taken 2009~August~10, while the lower panel is a radial brightness profile of the same mosaic.  \label{iawave_profile}}
\end{center}
\end{figure*}

\section{Spatial Context \label{Context}}

The two most prominent components of Saturn's ring system are the A~ring and the B~ring.  They are separated by the Cassini Division, named for its 17th-century discoverer, which turns out to be a multifaceted ring region in its own right, similar in character to the inner and equally tenuous C~ring (\Fig{}~\ref{ContextFig1}).  Surface densities in the main part of the Cassini Division are 1~to 3~g~cm$^{-2}$ \citep{soirings,Colwell09}, much lower than the characteristic values of 40~g~cm$^{-2}$ for the A~ring \citep[e.g.,][]{soirings} and the even higher values characteristic of the B~ring \citep{Robbins10}.  

The outermost part of the Cassini Division is the ``Ramp,'' so called because both the optical depth $\tau$ and the apparent brightness\fn{$I/F$ is a measure of observed brightness normalized by the incident solar flux density.} $I/F$ in this region increase gradually with ring radius from the low values of the main part of the Cassini Division, reaching relatively higher values at the inner edge of the A~ring, at which both parameters then jump to much higher values (\Fig{}~\ref{ContextFig2}).  Infrared spectroscopy indicates that the composition of material in the Ramp has more in common with the A~ring than it does with the main part of the Cassini Division \citep{HedmanSpectra13}.  A very similar Ramp marks the outermost regions of the C~ring, with $\tau$ and $I/F$ rising with ring radius toward the inner edge of the B~ring \citep{ColwellChapter09}. 

The inner edge of the A~ring is traditionally marked at a sharp jump in both $\tau$ and $I/F$ occurring at 122,050~km from Saturn's center \citep[\Fig{}~\ref{ContextFig2}; see also][]{French93}.  It has heretofore been presumed that there is a correspondingly sharp jump in surface density $\sigma$ at the same location, as that is the most straightforward explanation for a higher $\tau$.  Although there is no clear reason why there should be a boundary of any kind at this location,\fn{One recent suggestion is that the Mimas~2:1 resonance, which now forms the outer edge of the B~ring (inner edge of the Cassini Division) was once at this location, and that subsequent inward orbital migration of Mimas excavated what is now the Cassini Division \citep{LaineyDPS12,Lainey12}.} it was suggested by \citet{Durisen92} that, once formed, any boundary would be sharpened via ballistic transport due to bombardment of the ring by interplanetary meteoroids. 

A few hundred km outward of the A~ring's inner edge is a second sharp jump in both $\tau$ and $I/F$, this one at the location of the Pandora 5:4 Lindblad resonance at 122,313~km, which drives a strong density wave.  Outward of this resonant location is a region several thousand~km in annular width, characterized by seemingly chaotic radial structure (\Fig{}~\ref{ContextFig2}) that may be due to viscous overstability \citep{Thomson07,HedmanDPS12,RL13}, a process by which an over-active restoring force causes small perturbations in density to grow into concentric radial structure with wavelengths $\sim$100~m \citep{SchmidtChapter09}.  The Janus/Epimetheus~4:3 resonance at $\sim$125,250~km marks the outer boundary of the chaotic/overstable region, beyond which is the relatively quiescent mid-A~ring. 

As shown in \Fig{}~\ref{iawave_profile}, the appearance of this region is greatly altered in the equinox images.  The optical depth structure that usually dominates the apparent brightnes (\Fig{}~\ref{ContextFig2}) is greatly washed out due to the edge-on illumination, and the apparent brightness is instead dominated by a quasi-periodic pattern generated by the vertical corrugation of the bending wave (illustrated in \Fig{}~\ref{BendingWaveCartoon}).  

\section{Resonances and Spiral Waves \label{SpiralWaves}}
\subsection{General Introduction}
Spiral waves \citep{GT82,Shu84} occur in planetary rings at locations of resonance between an external forcing frequency, usually a moon's mean motion $n'$ (that is, its average angular velocity), and the natural orbital frequencies of ring particles.  The most widespread understood structure in Saturn's rings are \textit{spiral density waves} \citep[see][and references therein]{soirings}, in which the eccentricities of ring particles on resonant orbits are pumped up, leading to a compression wave in which both particle motions and wave propagation are radial.\fn{Particle motions are also excited in the azimuthal direction for both types of waves, which partly defines the waves' spiral structure; and, of course, the unexcited (keplerian) particle motions are azimuthal.}  Also significant are \textit{spiral bending waves} \citep[see, e.g.,][]{Shu83}, in which it is the inclinations of resonant ring particles that are pumped up, leading to a transverse wave that also propagates radially but with particle motions in the vertical direction. 

Spiral waves occur at locations where the ring particle mean motion $n$ is near an integer ratio (such as 3:2, 5:3, or 10:9) with the perturbing moon's mean motion $n'$.  This is formally expressed in terms of the resonance argument $\varphi$, which librates about a constant value (i.e., $\dot{\varphi} \approx 0$) for particles in the resonance.  The most prominent class of resonances in Saturn's rings are first-order (i.e., the two integers differ only by one), with the rings interior to the perturbing moon.  For a spiral density wave of this type, $\dot{\varphi}$ is of the form 
\begin{equation}
\dot{\varphi}_\mathrm{DW} = m n' - (m - 1) n - \dot{\varpi} , 
\label{ResArgDW}
\end{equation}
where $m$ is an integer identifying\fn{$m$ is also the number of spiral arms in the resulting density/bending wave.} the $m$:($m$-1) resonance, and $\dot{\varpi}$ is the ring particle's apsidal precession rate.\fn{The line of apsides is the long axis of an orbital ellipse, assuming non-zero eccentricity; thus, \textit{apsidal precession} is the angular motion of the orbital ellipse within the orbit plane.  The line of nodes is the line of intersection between the orbit plane and the reference plane, when the inclination is non-zero; thus, \textit{nodal precession} is the angular motion of the orbit plane.}  The resonance argument for spiral bending waves is similar, but with the ring particle's nodal precession rate $\dot{\Omega}$ instead of $\dot{\varpi}$.  The precession rates are non-zero because Saturn's asphericity, whose effect on the gravity field is described to first order by the parameter $J_2$, causes a ring particle's radial and vertical frequencies to differ from its mean motion \citep[see, e.g.,][]{MD99}.  However, there is an unusual class of spiral waves for which $n$ does not appear in the relevant equations; these are driven by commensurability between the perturbing moon's mean motion and the ring particle's precession alone.  

The simplest of these is an \textit{apsidal density wave}, driven by a simple commensurability between the perturber's mean motion $n'$ and the ring particle's apsidal precession rate $\dot{\varpi}$, which is to say that the long axis of the ring particle's elliptical orbit points toward the perturbing moon.\fn{This is identical to the \textit{evection resonance} that appears in the dynamics of Earth's Moon and other satellites, in that case with the Sun playing the part of the perturber.}  The resonance argument is simply
\begin{equation}
\dot{\varphi}_\mathrm{ADW} = n' - \dot{\varpi} ,
\label{ResArgADW}
\end{equation}
which naturally leads to the resonance being labeled ``1:0,'' after the respective integer coefficients of $n'$ and $n$, with reference to \Eqn{}~\ref{ResArgDW}.  A similar commensurability involving $n'$ and the ring particle nodal precession rate $\dot{\Omega}$ leads to the \textit{nodal bending wave} \citep{RL88}, in which it is the ring particle's orbit plane that precesses in time with the perturber's mean motion.  However, for an oblate planet such as Saturn, nodes precess ``backwards'' (i.e. $\dot{\Omega}<0$); thus,\fn{Another reason for the more complex resonance argument for the nodal bending wave is one of d'Alembert's rules \citep[see, e.g.,][]{Ham94,MD99}, which states that the sum of all nodal coefficients in a valid resonance argument must be an even number (in \Eqn{}~\ref{ResArgNBW}, the coefficients of $\dot{\Omega}'$ and $\dot{\Omega}$ sum to zero, satisfying the rule).  For mathematical details of nodal resonances, see \citet{RL88}.} the simplest argument that constitutes a valid resonance is
\begin{equation}
\dot{\varphi}_\mathrm{NBW} = - n' + \dot{\varpi}' + \dot{\Omega}' - \dot{\Omega} .
\label{ResArgNBW}
\end{equation}
While interpreting \Eqn{}~\ref{ResArgNBW}, keep in mind that the middle two terms are very small compared to the outer two, since apsidal and nodal resonances involve perturbing moons that are so far from the planet that their orbital rates are comparable to the precession rates of ring particle orbits, but that both $\dot{\Omega}$ and $\dot{\Omega}'$ are negative.  Again following (one might say \textit{ad absurdam}) the system for labeling resonances and their spiral waves based on the integer coefficients of $n'$ and $n$, the nodal resonance is labeled ``-1:0.''

Because of the ``backwardness'' inherent in the nodal resonance, nodal bending waves propagate radially outward (when the perturber is outward of the rings, as is the case for all resonances discussed in this work) as do density waves, rather than inward as do other bending waves. Furthermore, because the frequencies involved with apsidal and nodal resonances are much lower than for other wave-producing resonances (due to the absence of the ring particle's mean motion in $\dot{\varphi}$), their waves have much longer wavelengths than typical spiral waves and are excited by moons orbiting much farther from Saturn.  Indeed, while typical spiral waves are excited by the inner moons Pan, Atlas, Prometheus, Pandora, Janus, Epimetheus, and Mimas, the only known moons whose apsidal and nodal resonances fall within the main rings are Titan, Hyperion, and Iapetus.  Of these, Titan's apsidal resonance governs an eccentric ringlet in the mid-C~ring \citep{Porco84a,NP88} while its nodal bending wave is nearby \citep{RL88}, Hyperion's waves are weak but have been seen in the outer-C~ring  
and will be discussed in another publication by M.~S.~Tiscareno and B.~E.~Harris (in preparation), and Iapetus' waves are the subject of this work. 

\subsection{Dispersion Relations and Wavenumber Profiles \label{Wavenumber}}
Because their low amplitudes fall far short of saturating the background, all spiral waves considered in this work follow the linear theory \citep{GT82,Shu84}, relevant parts of which we now reproduce.  Mindful of some variations among authors, we largely follow the notation of \citet{Rosen91a}, as \citet{soirings} did with more comment.  

The dispersion relation for an apsidal spiral density wave is 
\begin{equation}
\label{DispersionADW}
[n' - n(r)]^2 = [n(r) - \dot{\varpi}(r)]^2 - 2 \pi G \sigma_\mathrm{b} | k(r) | ,
\end{equation}
where $G$ is Newton's constant, $\sigma_\mathrm{b}$ is the background surface density, and $n$ and $\dot{\varpi}$ and the wavenumber $k=2 \pi/\lambda$ (where $\lambda$ is the spatial wavelength) are functions of the radial location $r$.  Similarly, the dispersion relation for a nodal spiral bending wave is
\begin{equation}
\label{DispersionNBW}
[- n' + \dot{\varpi}' + \dot{\Omega}' - n(r)]^2 = [n(r) - \dot{\Omega}(r)]^2 - 2 \pi G \sigma_\mathrm{b} | k(r) | . 
\end{equation}
Both these dispersion relations can be linearized to
\begin{equation}
\label{WavenumberEqn1}
|k(r)| = \frac{|\mathscr{D}| |r-r_\mathrm{res}|}{2 \pi G \sigma_\mathrm{b} r_\mathrm{res}} , 
\end{equation}
where $r_\mathrm{res}$ is the radial location of the resonance and 
\begin{equation}
\label{ScriptyD}
\mathscr{D} = \frac{21}{2} J_2 \left(\frac{R_\mathrm{S}}{r_\mathrm{res}}\right)^2 n^2(r_\mathrm{res})
\end{equation}
for $m=1$ resonances \citep{Cuzzi84}.  Since both $\mathscr{D}$ and $r-r_\mathrm{res}$ are always positive in the $m=1$ case, we can combine \Eqn{s}~\ref{WavenumberEqn1} and~\ref{ScriptyD} to write the 
wavenumber as
\begin{equation}
\label{WavenumberEqn2}
k(r) = \frac{21 J_2 R_\mathrm{S}^2 M_\mathrm{S}}{4 \pi r_\mathrm{res}^6} \cdot \frac{r-r_\mathrm{res}}{\sigma_\mathrm{b}} . 
\end{equation}
Using standard values for Saturn of $R_\mathrm{S} = 60,$330~km, $M_\mathrm{S} = 5.68 \times 10^{26}$~kg, and $J_2 = 0.01629$ \citep{Jake06}, we can write the wavelength $\lambda \equiv 2 \pi / k$ for $m=1$ resonances in the useful form 
\begin{equation}
\label{WavenumberEqn3}
k(r) = \left( 0.0117 \mathrm{~km}^{-2} \right) \left( r-r_\mathrm{res} \right) \left(\frac{R_\mathrm{S}}{r_\mathrm{res}}\right)^6 \left( \frac{10 \mathrm{~g~cm}^{-2}}{\sigma_\mathrm{b}} \right) . 
\end{equation}

When the background surface density $\sigma_\mathrm{b}$ is constant, it is evident from \Eqn{}~\ref{WavenumberEqn3} that $\ud k / \ud r$ is constant.  Most spiral waves in Saturn's rings span a small enough region that this assumption is robust, such that $\sigma_\mathrm{b}$ can be obtained by finding the slope of a line fit to $k(r)$, as has been done by several workers with \Cassit{} data \citep{soirings,Colwell09,Baillie11}.  Some of the complex waves due to the co-orbital moons Janus and Epimetheus were fit using a different method by \citet{jemodelshort}.  The waves discussed in this work, on the other hand, span large regions of this disk, over which the background surface density varies widely.  In this case, as a consequence of the WKB approximation that underlies linear spiral wave theory \citep{GT82,Shu84}, we can still use \Eqn{}~\ref{WavenumberEqn3} with the measured wavenumber $k(r)$ at each radial location giving the local surface density $\sigma_\mathrm{b}(r)$. 

\section{Observations \label{Observations}}

The Iapetus nodal bending wave is visible in \Cassit{}~ISS images taken during an interval broadly extending from 2008 to~2010, when the solar incidence angle $\iota$ was within several degrees of $90^\circ$ (though it was not noticed until it became very prominent in images taken in spring~2009).  The feature is more prominent as $\iota$ approaches $90^\circ$ (equinox), so we concentrate our analysis on two images taken during Orbit~116, with portions of the Sun's disk on both sides of the ring plane.  See Table~\ref{iawave_imtable} for relevant geometrical parameters.  Due to practical constraints on the viewing geometry, these images were taken at a local longitude of $109.3^\circ$ while the Sun was at a longitude of $224.5^\circ$.  The difference between the two longitudes is $-115.2^\circ$, closer to local morning than to local noon or midnight, the latter of which are more optimal for highlighting vertical structure (\Fig{}~\ref{BendingWaveCartoon}), yet far enough from local morning that the vertical structure is sufficiently highlighted.  

\begin{table}[!b]
\caption{Observing information for images used in this work. \label{iawave_imtable}}
\begin{scriptsize}
\begin{tabular} { c c c c c c c c c }
\hline
\hline
 & & & Incidence & Emission & Phase & Radial & Azimuthal & Exposure \\
Image & Orbit & Date/Time & Angle$^a$ & Angle$^a$ & Angle & Resolution$^b$ & Resolution$^b$ & Duration$^c$ \\
\hline
N1560311316 & 046 & 2007-163 03:14:24 & 102.06$^\circ$ & 125.4$^\circ$ &  44.1$^\circ$ &   0.9 &    0.9 &   0.5 \\
N1560311433 & 046 & 2007-163 03:16:21 & 102.06$^\circ$ & 125.0$^\circ$ &  43.8$^\circ$ &   0.9 &    0.9 &   0.5 \\
N1560311549 & 046 & 2007-163 03:18:17 & 102.06$^\circ$ & 124.5$^\circ$ &  43.6$^\circ$ &   1.0 &    0.9 &   0.5 \\
\hline  
N1628594653 & 116 & 2009-222 10:42:04 &  90.01$^\circ$ &  94.8$^\circ$ & 150.0$^\circ$ &   7.0 &   45.1 &   1.0 \\
N1628594713 & 116 & 2009-222 10:43:04 &  90.01$^\circ$ &  94.8$^\circ$ & 150.1$^\circ$ &   7.0 &   45.1 &   0.7 \\
\hline  
\multicolumn{9}{l}{$^a$ Measured from the direction of Saturn's north pole (ring-plane normal), so that $90^\circ$ denotes edge-on and $>$90$^\circ$ denotes} \\
\multicolumn{9}{l}{the southern hemisphere.  Note that the Sun's angular size, as seen from Saturn, is 0.06$^\circ$.}\\
\multicolumn{9}{l}{$^b$ In km/pixel.}\\
\multicolumn{9}{l}{$^c$ In seconds.}\\
\end{tabular}
\end{scriptsize}
\end{table}

All images were calibrated and converted to values of $I/F$ using the standard Cisscal package v3.6 \citep{PorcoSSR04}.  The images were navigated using stars and ring edges as fiducials.  We then used SPICE geometry software \citep{Spice} with NAIF kernels\fn{Available at \texttt{ftp://naif.jpl.nasa.gov}} encoding the motions of Saturn, its moons, and the spacecraft, to assign a ring radius (i.e., distance from Saturn center within the ring plane) to each pixel and co-add pixels appropriately to assemble a radial brightness profile for each image.  Further conversion could be attempted, from simple observed brightness to the ring's optical depth, and from optical depth to surface density.  However, these are difficult because of the complex ways in which ring material interacts with light \citep[e.g.,][]{Anparsgw10}, especially during equinox; furthermore, they are unnecessary because we are interested only in the spatial periodicities of the waves we are examining, and those periodicities can be inferred from the brightness profiles alone since the optical depth and surface density are known to vary with brightness in a relatively simple manner \citep{Cuzzi84}.  

\begin{figure*}[!t]
\begin{center}
\includegraphics[width=16cm]{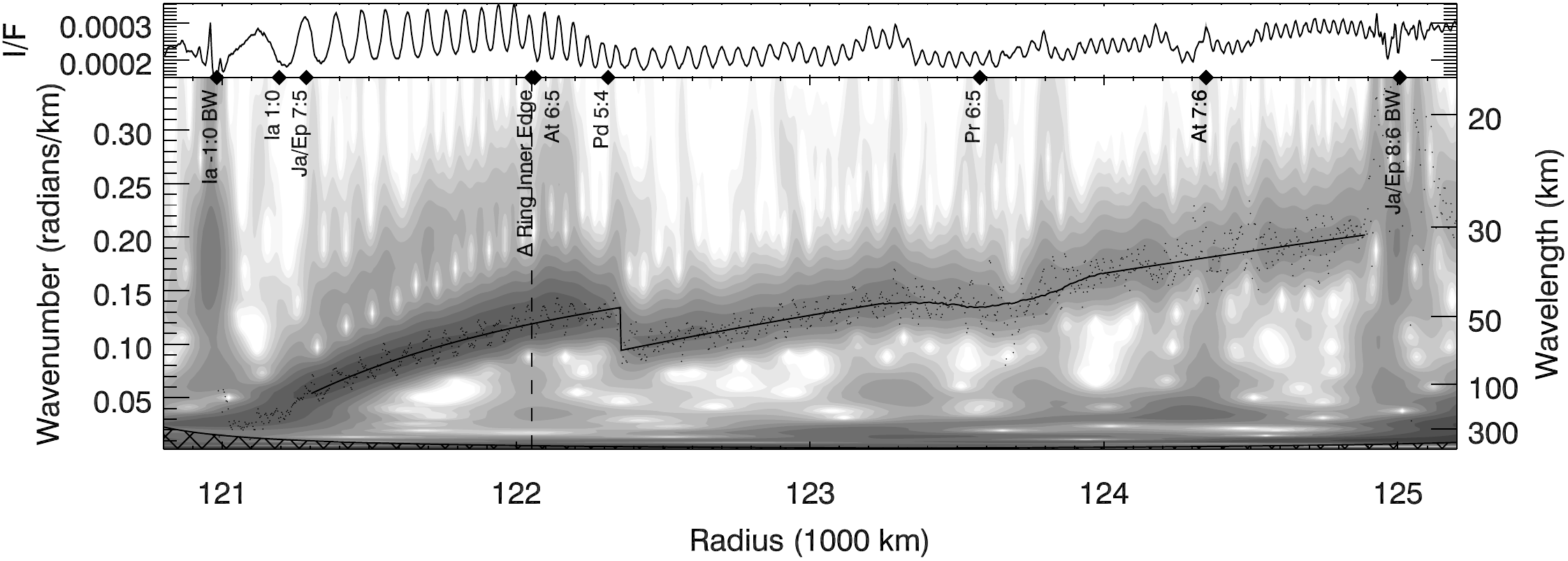}
\caption{The upper panel shows the radial brightness profile of the region of interest taken during equinox (from images N1628594653 and N1628594713, taken on 2009~August~10), the same as in the lower panel of \Fig{}~\ref{iawave_profile}.  The lower panel shows the Morlet wavelet transform of that brightness profile.  Resonances and other features of interest are shown as filled circles on the line between the panels and labeled with text; note the $\sim$200-km radial separation between the locations of the Iapetus~-1:0 nodal resonance and the Iapetus~1:0 apsidal resonance.  The canonical inner edge of the A~ring is also shown as a vertical dashed line.  The solid line is a fitted wavenumber profile for the Iapetus~-1:0 nodal bending wave, corresponding to the surface density profile shown in \Fig{}~\ref{iawave_sigmamodel}.  The scattered dots are the raw data from which the fitted profile was derived (see \Fig{}~\ref{iawave_protosigmamodel}). 
\label{iawave_wavelet}}
\end{center}
\end{figure*}

\begin{figure*}[!t]
\begin{center}
\hspace{-1.5cm}
\includegraphics[width=15cm]{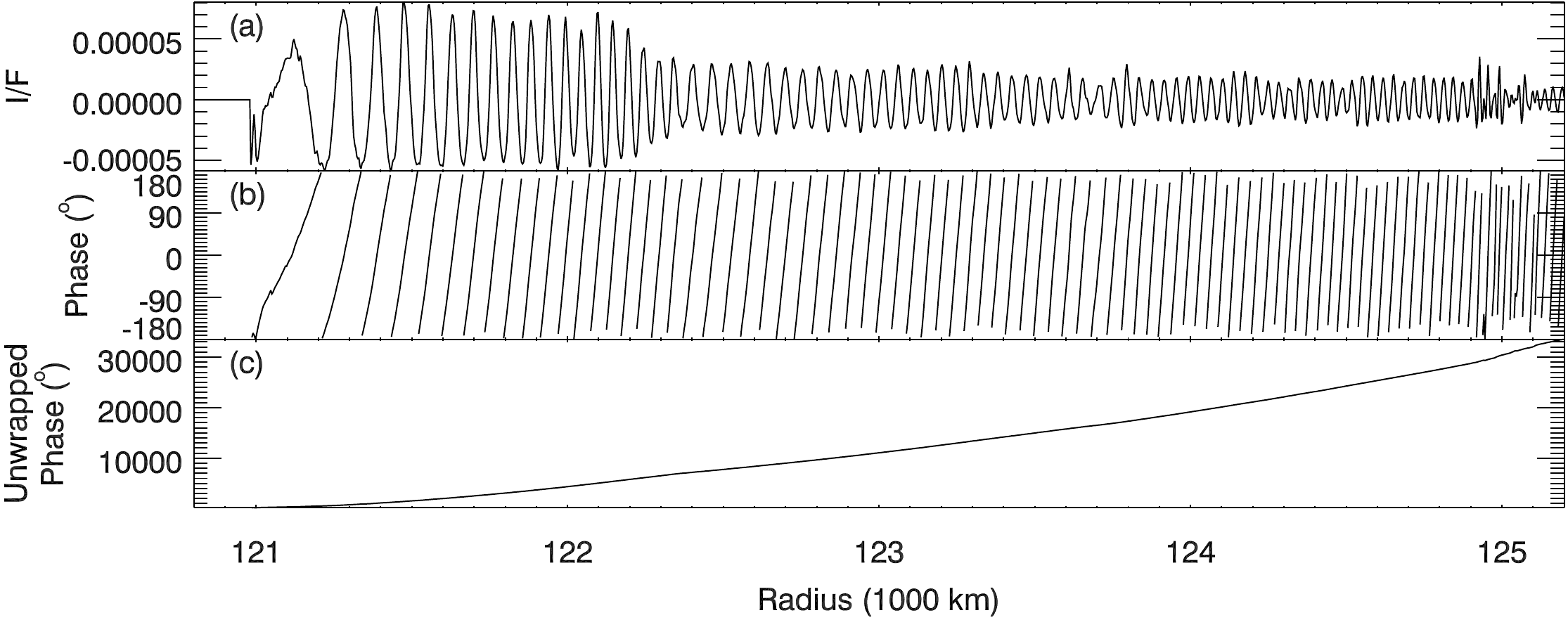}
\caption{(a) High-pass filtered (see Section~\ref{Iapetus10Analysis}) radial brightness profile of the region of interest taken during equinox (for unfiltered version, see lower panel of \Fig{}~\ref{iawave_profile} and upper panel of \Fig{}~\ref{iawave_wavelet}).  (b) Wavelet phase $\bar{\phi}_\mathrm{W}(r)$, which is zero near local peaks and 180$^\circ$ near local troughs.  (c) Unwrapped wavelet phase, compiled by simply adding to $\bar{\phi}_\mathrm{W}(r)$, which constantly increases, another factor of 360$^\circ$ every time it passes through 180$^\circ$.  
\label{iawave_waveletphase}}
\end{center}
\end{figure*}

\begin{figure*}[!t]
\begin{center}
\includegraphics[width=8cm]{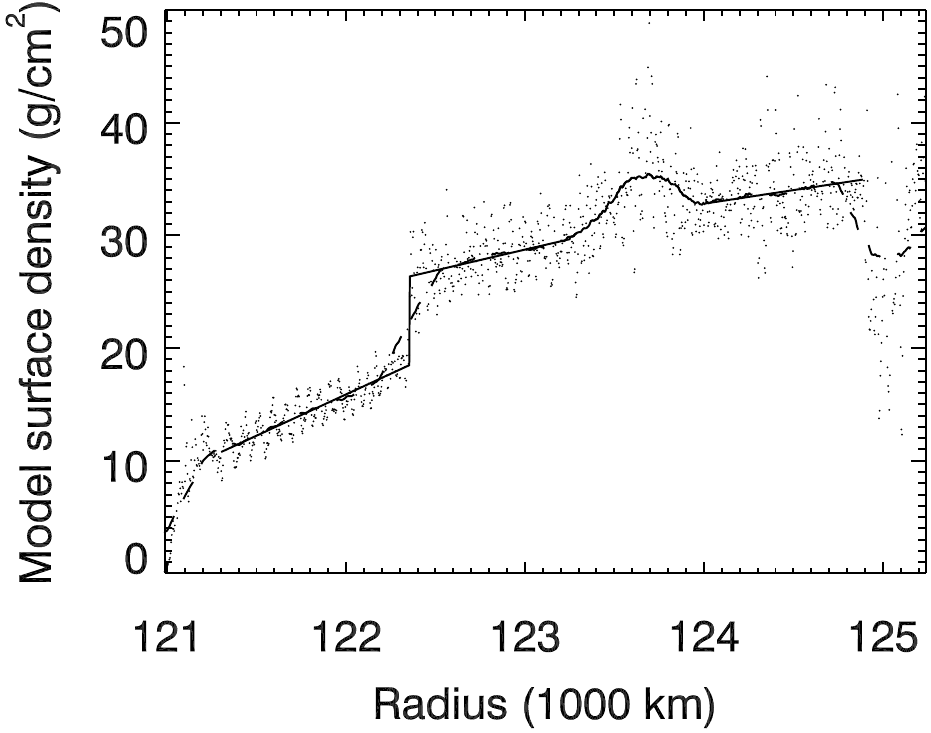}
\caption{The scattered dots are the calculated surface density $\sigma_\mathrm{b}$ at each radial location, calcualted from \Eqn{}~\ref{WavenumberEqn3} with $k = \ud \bar{\phi}_\mathrm{W} / \ud r$, where $\bar{\phi}_\mathrm{W}(r)$ is plotted in \Fig{}~\ref{iawave_waveletphase}c.  The dashed line is the result of a simple low-pass smoothing filter with a boxcar width of 100 elements.  The solid line is a more sophisticated model of the surface density profile, derived by taking piecewise linear fits on the intervals [121300,122357]~km, [122357,123250]~km, and [124000,124890]~km, and simply following the boxcar filter on the interval [123250,124000]~km.  The derived model follows the simple boxcar filter closely at all points except where the latter clearly fails to represent the data, at either end due to the inclusion of non-bending-wave points into the smoothing, and near 122357~km due to smoothing over the sharp discontinuity. 
\label{iawave_protosigmamodel}}
\end{center}
\end{figure*}

\section{Wavelet Analysis \label{Analysis}}

A Morlet wavelet transform, which acts as a spatially-resolved Fourier transform for quasi-sinusoidal signals with spatially-varying frequencies, was carried out on each radial brightness profile using techniques described by \citet{soirings}.  In the resulting periodograms, the horizontal axis is the radial coordinate $r$ while the vertical axis is the spatial wavenumber $k = 2 \pi / \lambda$ (where $\lambda$ is the spatial wavelength) and shading indicates the strength of a given frequency at a given radial location.  

\subsection{Iapetus -1:0 nodal bending wave \label{Iapetus10Analysis}}

The wavelet transform of the Iapetus~-1:0 nodal bending wave, as seen in greatest detail on 2009~August~10, is shown in \Fig{}~\ref{iawave_wavelet}.  It is clear that this feature is the -1:0 nodal bending wave, rather than the 1:0 apsidal density wave as claimed by \citet{Cuzzi81}, not only because of the strong dependence of its visibility on the solar incidence angle, but also because \Fig{}~\ref{iawave_wavelet} shows the first full wavelength of the outward-propagating wave, as well as much power in the wavelet transform, occuring inward of the Iapetus~1:0 apsidal resonance.  It should be noted that the \Voyit{} imaging data used by \citet{Cuzzi81} may have had insufficient spatial resolution to clearly make this distinction, and also that they wrote before \citet{RL88} first worked out the theory of nodal bending waves. 

A harmonic signal can be seen in the wavelet transform (\Fig{}~\ref{iawave_wavelet}) between 121,300~km and 122,200~km, with wavenumbers exactly twice that of the main signal.  This is a numerical artifact that arises because the signal is not perfectly sinusoidal \citep{soirings}. 

As described by \citet{soirings}, the most accurate way to track the changing wavenumber of a quasi-sinusoidal signal is to analyze the wavelet phase $\bar{\phi}_\mathrm{W}(r)$ --- which essentially measures the respective proximity of local peaks and troughs --- at each location, after first zeroing out the wavelet transform for wavenumbers lower than the desired signal (\Fig{}~\ref{iawave_waveletphase}a).  The latter is a form of high-pass filter that removes unwanted long-wavelength structure.  The wavelet phase is ``unwrapped'' to yield a monotonically increasing function of radius (\Fig{}~\ref{iawave_waveletphase}b), and the wavenumber is then calculated as $k = \ud \bar{\phi}_\mathrm{W} / \ud r$ and converted to the local surface density $\sigma_\mathrm{b}$ by means of \Eqn{}~\ref{WavenumberEqn3}.  

\begin{figure*}[!t]
\begin{center}
\includegraphics[width=8cm]{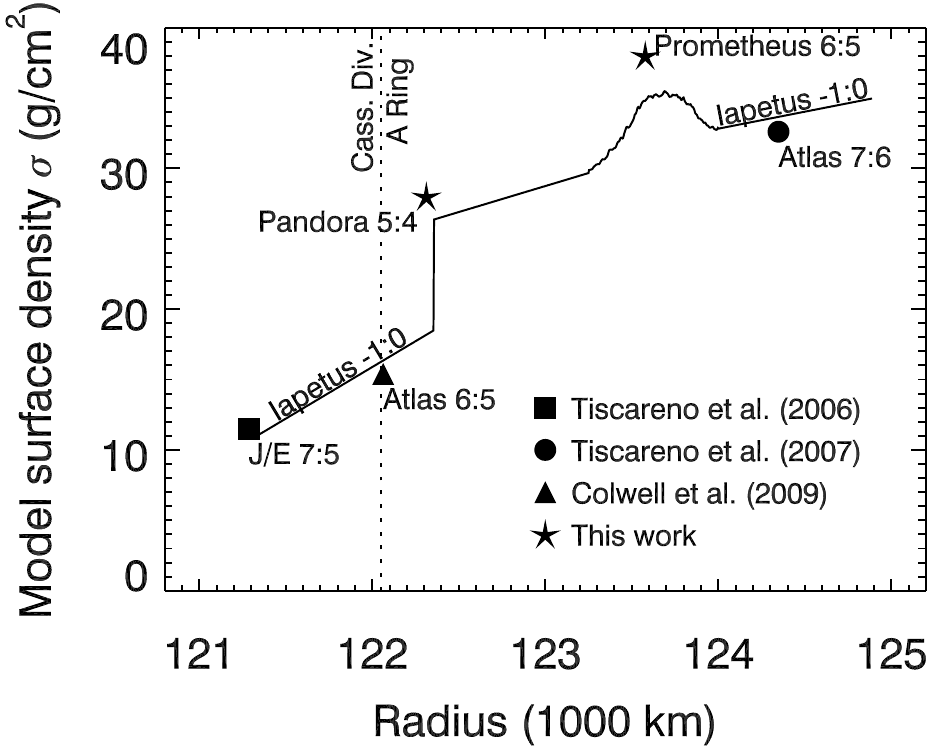}
\caption{The solid line denotes our derived surface density profile from the Iapetus~-1:0 nodal bending wave (\Fig{}~\ref{iawave_protosigmamodel}).  There is no sharp change in surface density $\sigma_\mathrm{b}$ to correspond with the observed sharp change in optical depth $\tau$ at the canonical inner edge of the A~ring (vertical dotted line; cf.~\Fig{}~\ref{ContextFig2}).  Other plotted symbols indicate independent measurements of the surface density from other (spatially smaller) spiral density waves.  
\label{iawave_sigmamodel}}
\end{center}
\end{figure*}

\begin{figure*}[!t]
\begin{center}
\includegraphics[width=9cm]{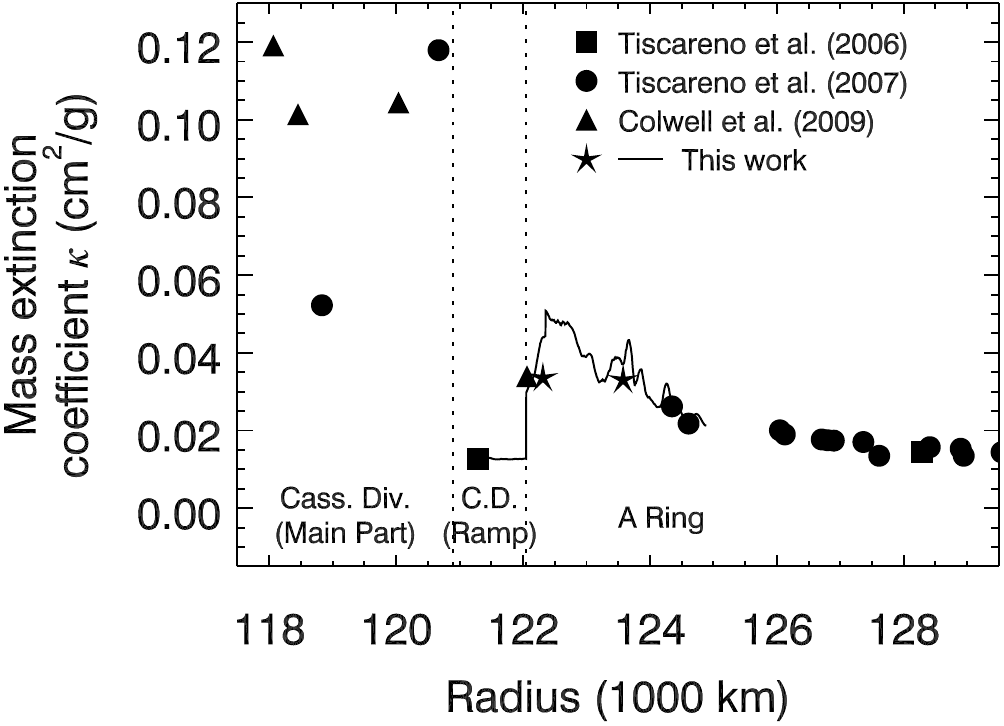}
\caption{Mass extinction coefficient $\kappa \equiv \tau/\sigma$, derived from the surface density $\sigma$ measured for individual spiral density waves and the background optical depth $\tau$ at each location (the values of $\tau$ are taken from the \Cassit{}~VIMS stellar occultation plotted in the lower panel of \Fig{}~\ref{ContextFig2}).  
The mass extinction coefficient appears to be highest in the main part of the Cassini Division, lowest in the Cassini Division Ramp, and to take an intermediate value in the inner-A~ring, trending gently lower into the mid-A~ring.  Overplotted as a solid line is the mass extinction coefficient derived by this work from the Iapetus~-1:0 nodal bending wave. 
\label{mecfig}}
\end{center}
\end{figure*}

The resulting estimates of local surface density $\sigma_\mathrm{b}(r)$ are shown in \Fig{}~\ref{iawave_protosigmamodel}.  The local scatter is due primarily to residual oscillations in $\bar{\phi}_\mathrm{W}(r)$ that arise because the signal is not perfectly sinusoidal, as well as gaussian variation in $\bar{\phi}_\mathrm{W}(r)$, both of which are amplified by the derivative used to calculate $k(r)$ and then $\sigma_\mathrm{b}(r)$.  In order to convert these data into a model that can be used for further analysis, smoothing out the scatter while preserving the sharp discontinuity at the Pandora~5:4 resonance location, we fit the data by several piecewise linear fits, as described in \Fig{}~\ref{iawave_protosigmamodel}.  The close correspondence between our derived model and a simple smoothing of the data, except at locations where the latter is clearly deficient, supports our derivation.  

Our derived $\sigma_\mathrm{b}(r)$ is plotted in \Fig{}~\ref{iawave_sigmamodel} along with independent measurements of $\sigma_\mathrm{b}$ at certain locations as ascertained by means of other spiral density waves\fn{Surface densities derived from the Pandora~5:4 and Prometheus~6:5 are reported for the first time in this work.  Starting with radial brightness scans of images N1560311549 and N1560311316, respectively, we simply found by eye the value of $\sigma_0$ that gave the best match to the slope of the wave's wavelet signature in radius-wavenumber space (see \Eqn{}~\ref{WavenumberEqn1}, which applies for all values of $m$), as irregular waveforms (see \Fig{}~\ref{Pandora54}) preclude use of the more sophisticated method developed by \citet{soirings}.  We obtained $\sigma_0 = 28 \pm 5$~g~cm$^{-2}$ from the Pandora~5:4 wave originating at $r = 122$,313~km, and $\sigma_0 = 38 \pm 5$~g~cm$^{-2}$ from the Prometheus~6:5 wave originating at $r = 123$,578~km.} (keep in mind that the nodal bending wave has much longer wavelengths and covers much more territory than common spiral waves, as described in Section~\ref{SpiralWaves}).  The agreement gives further support to our derived model.  Additionally, the wavenumber $k(r)$ as derived from our model $\sigma_\mathrm{b}(r)$ by means of \Eqn{}~\ref{WavenumberEqn3} is overplotted on the wavelet transform in \Fig{}~\ref{iawave_wavelet}.  

The mass extinction coefficient $\kappa \equiv \tau/\sigma$ measures the ring's light-blocking ability per unit surface density.  Higher values of $\kappa$, indicating more light-blocking surface area for a given surface density, are most simply achieved by breaking the mass into smaller particles.  Contrariwise, lower values of $\kappa$ are most simply achieved by concentrating the mass into large particles or clumps.  The mass extinction coefficient as derived by this work is placed in the context of previous observations in \Fig{}~\ref{mecfig} and clarifies the situation considerably.  It is now evident that the Cassini Division Ramp is a region of particularly low $\kappa$, as much as 10x lower than in the main part of the Cassini Division, while values of $\kappa$ in the inner-A~ring are intermediate and trend gently lower into the mid-A~ring.  

\subsection{Pandora 5:4 density wave \label{Pandora54Analysis}}

\begin{figure*}[!t]
\begin{center}
\includegraphics[width=16cm]{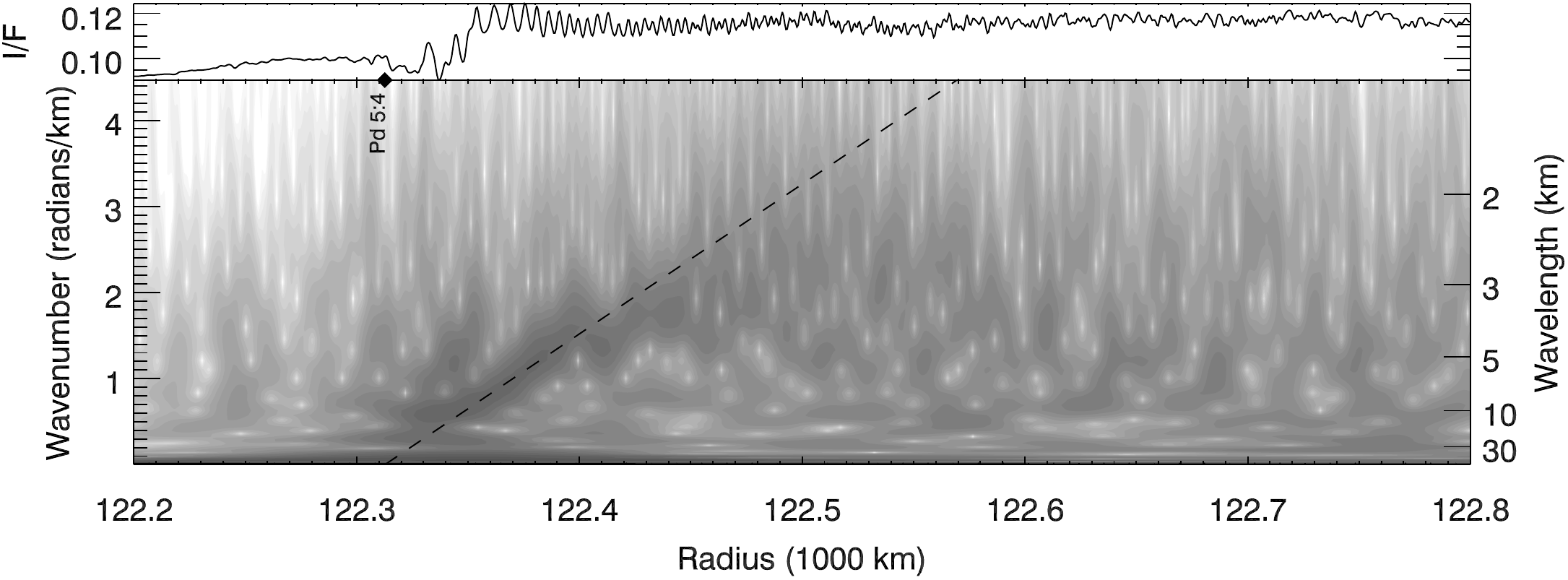}
\caption{The upper panel shows a radial brightness profile of the region surrounding the Pandora~5:4 density wave, taken with its normal appearance (the Sun was 12$^\circ$ out of the ring plane; see Table~\ref{iawave_imtable}).  The images are N1560311433 and N1560311549, taken on 2007~June~12.  As in \Fig{}~\ref{iawave_wavelet}, the lower panel shows the Morlet wavelet transform of that brightness profile.  The location of the Pandora~5:4 Lindblad resonance is marked at 122,313~km, and the dashed line shows the expected wavenumber trace for $\sigma_\mathrm{b} = 28$~g~cm$^{-2}$ (\Eqn{}~\ref{WavenumberEqn3}), which fits the data.  
\label{Pandora54}}
\end{center}
\end{figure*}

While it has been commonly repeated that the Pandora~5:4 density wave has the longest train of any in the A~ring \citep[e.g.,][]{ColwellChapter09}, in fact it can be seen in \Fig{}~\ref{Pandora54} that only the inner $\sim$100~km of the oscillations have significant power in the $k \propto (r-r_\mathrm{res})$ pattern that is characteristic of propagating density waves.  The remaining several thousand~km of high-frequency structure does not show any evidence of frequencies related to the Pandora~5:4 density wave.  Furthermore, given the surface density model shown in \Fig{}~\ref{Pandora54}, the wavelength of the Pandora~5:4 wave would reach values as small as  
tens of meters, if indeed it propagated all the way to the outer edge of the high-frequency region at 125,250~km.  

Rather, it is likely that the Pandora~5:4 wave merely serves as the inner boundary of a region marked by high spatial frequencies (as well as high surface densities) that are likely due to viscous overstability \citep{Thomson07,Colwell07,SchmidtChapter09} and perhaps other processes.  Whether this spatial arrangement occurs through causality or simply by coincidence is not clear.  Because Lindblad resonances transfer negative angular momentum into the disk, it should not be possible for the Pandora~5:4 wave to actively maintain the mass shelf at its location (as, for example, the Mimas~2:1 and Janus/Epimetheus~7:6 resonances respectively maintain the \textit{outer} edges of the B~and A~rings).  On the other hand, the strong oscillations of the Pandora~5:4 density wave may play a role in triggering the viscous overstability in a region that, for other reasons, is susceptible to it. 

\section{Discussion \label{Discussion}}

The lack of a sharp jump in surface density at the inner edge of the A~ring (\Fig{}~\ref{iawave_sigmamodel}) is surprising and difficult to explain.  What could drive the observed sharp jump in optical depth, if not the surface density?  In other words, what could drive a sharp jump in the mass extinction coefficient~$\kappa$ at that location (\Fig{}~\ref{mecfig})?  

One way for the optical depth to change sharply with only a modest trend in surface density is if the surface density were to pass through a threshold value at which self-gravity wakes (SGWs) ``turned on.''  Indeed, SGWs do exist throughout the inner A~ring but not in the Cassini Division Ramp \citep{NH10}.  SGWs, which are pervasive clumps due to a gravitational instability that is not strong enough to run completely to accretion \citep[e.g.,][]{SchmidtChapter09} can indeed have a sudden onset as well as a strong effect on the mass extinction coefficient \citep{Daisaka01,Anparsgw10}.  However, the observed effect is in the wrong direction, as the onset of SGWs would be triggered by rising surface density and would \textit{decrease} the optical depth for a given surface density.  Therefore, SGWs cannot answer the puzzle posed by our findings. 

The most straightforward interpretation of these observations is that the inner-A~ring is composed of smaller particles (on average) than the Cassini Division Ramp, while the main part of the Cassini Division is composed of still smaller particles.  Why the Cassini Division Ramp would have such large particles, and why there would be a sharp boundary in the particle-size distribution, is not clear.  Ballistic transport \citep{Durisen92} may conceivably operate to sharpen an edge in optical depth without an edge in surface density, or particles of different sizes may migrate radially at different rates in such a way as to give rise to a separation, but further study is needed to determine whether that is possible.  Another possibility is that the Cassini Division Ramp was the site of an impact that seeded that region with larger particles that have not yet come to collisional equilibrium, but this too requires further study to evaluate its plausibility.  Further study is also needed to determine whether the varying particle-size distributions suggested in this paragraph can be reconciled with the ``excess variance'' patterns measured by stellar occultations \citep{SN90,ColwellDPS11} and with the multi-wavelength diffraction data from \Cassit{} radio occultations \citep[e.g.,][]{CuzziChapter09}

Finally, our findings raise the question of whether the Cassini Division Ramp would more usefully be reckoned as the innermost part of the A~ring, rather than as the outermost part of the Cassini Division (similarly, whether the C~ring Ramp might have more in common with the inner-B~ring than with the rest of the C~ring).  A possible compositional affinity was previously mentioned in Section~\ref{Context}.  The densest part of the A~ring is the apparently ``chaotic region'' from 122,350~km to $\sim$125,000~km, which is likely characterized by viscous overstability.  The inner edge of that region is a precipitous drop in both optical depth and surface density at the location of the Pandora~5:4 resonance (122,313~km), but the boundaries are not sharp.\fn{In fact, the lack of a reflected wave propagating inward from this location (\Fig{}~\ref{iawave_wavelet}) indicates that the transition in surface density is not much sharper than the wavelength at that location, $\sim$50~km.}  Further inward (see \Fig{s}~\ref{ContextFig2} and~\ref{iawave_sigmamodel}) is a 250-km ``doorstep'' region of intermediate optical depth, whose inner boundary at the canonical inner edge of the A~ring (122,050~km) is now revealed as a sharp edge in optical depth but not in surface density (thus, as a sharp edge in mass extinction coefficient $\kappa$, most easily explained by means of particle properties).  The surface density decreases steadily inward from the Pandora resonance to the inner reaches of the Ramp.  Perhaps the real inner edge of the A~ring complex should be reckoned at 120,900~km, where the Cassini Division's ``triple band'' feature gives way to the first risings of the Ramp.  

However, no change in nomenclature will solve the primary question raised by this work:  What is the nature of the changes in ring properties at 122,050~km, the canonical inner edge of the A~ring, and why do they occur? 

\vspace{1cm}
\noindent \textbf{Acknowledgements} \\We thank Jeff Cuzzi and Peter Goldreich for helpful conversations, and J\"urgen Schmidt and Larry Esposito for improvements to the manuscript.  We thank the Cassini Project and the Cassini Imaging Team.  M.S.T. acknowledges funding from the NASA Cassini Data Analysis program (NNX08AQ72G and NNX10AG67G) and the Cassini Project.

\end{document}